\input{psfig.sty}

\documentclass{aa}

\def\saxj{SAX J1808.4--3658}
\def\aql{Aql X--1}

\begin{document}

   \thesaurus{06     % A&A Section 6: Form. struct. and evolut. of stars
              (03.11.1;  % Cosmogony,
               16.06.1;  % Planets and satellites: general,
               19.06.1;  % Solar system: general,
               19.37.1;  % Stars: formation of,
               19.53.1;  % Stars: oscillations of,
               19.63.1)} % Stars: structure of.
\title{Constraints on the Neutron Star Magnetic Field of the two X-ray 
Transients \saxj\ and \aql}

  \subtitle{}

   \author{Tiziana Di Salvo \inst{1} \and Luciano Burderi \inst{2}}
   
   \offprints{T. Di Salvo}

   \institute{Astronomical Institute "Anton Pannekoek," University of 
              Amsterdam and Center for High-Energy Astrophysics,
              Kruislaan 403, NL 1098 SJ Amsterdam, the Netherlands\\
              email: disalvo@astro.uva.nl
         \and
             Osservatorio Astronomico di Roma, Via Frascati 33, 
             00040 Monteporzio Catone (Roma), Italy \\
             email:  burderi@coma.mporzio.astro.it}

   \date{Received ; accepted }
   
   \authorrunning{T. Di Salvo \& L. Burderi}
   \titlerunning{On the neutron star magnetic field of \saxj\ and \aql}

   \maketitle

\begin{abstract}

The recently discovered coherent X-ray pulsations at a frequency of 
$\sim 400$ Hz in \saxj, together with a measure of the source luminosity 
in quiescence, allow us to put an upper limit on the neutron star magnetic 
field, that is $B \le 5 \times 10^8$ Gauss, using simple considerations on
the position of the magnetospheric radius during quiescent periods. Combined 
with the lower limit inferred from the presence of X-ray pulsations, 
this constrains the \saxj\ neutron star magnetic field in the quite
narrow range $(1-5) \times 10^8$ Gauss. Similar considerations applied to the
case of \aql\ give a neutron star magnetic field lower than $\sim 10^9$ Gauss.


\end{abstract}

\section{Introduction}

Low-mass X-ray binaries (LMXBs) consist of a neutron star, generally with a 
weak magnetic field ($B \la 10^{10}$ Gauss), accreting matter from a low-mass 
($\la 1\ M_\odot$) companion. Neutron star soft X-ray transients (hereafter
SXT) are a special subgroup of LMXBs. SXTs are usually found in a 
quiescent state, with luminosities in the range $10^{32}-10^{33}$ ergs/s.
On occasions they exhibit outbursts, during which the luminosity increases to 
$\sim 10^{36}-10^{38}$ ergs/s and their behavior closely resemble that of
persistent LMXBs. SXTs indeed form a rather inhomogeneous
class, with sources showing regular outbursts (e.g.\ Aql X--1, Cen X--4,
4U 1608--522) and sources with long on/off activity periods (e.g.\ KS
1731--260, X 1732--304; see Campana et al.\ 1998 for a review).  

The mechanism for the quiescent X-ray emission in these sources is still
uncertain (e.g., Menou et al. 1999; Campana \& Stella 2000; Bildsten \& 
Rutledge 2001).  The spectrum in quiescence is usually well fit by a soft
thermal component (blackbody temperature of $\sim 0.1-0.3$~keV) plus
a power-law component with a photon index $\Gamma \sim 1-2$.
The blackbody component is interpreted as thermal emission from a pure
hydrogen neutron-star atmosphere (e.g.\ Rutledge et al. 1999, 2000), while 
the power-law component is thought to be due to residual accretion or the 
interaction of a pulsar wind with matter released by the 
companion star (see e.g.\ Campana \& Stella 2000, and references therein). 

Some of these neutron star SXTs also show type-I X-ray bursts. 
During these bursts nearly-coherent oscillations are sometimes observed,
the frequencies of which are in the rather narrow range between 300 and 600~Hz 
(see van der Klis 2000; Strohmayer 2001 for reviews). This frequency is 
interpreted as the neutron star rotation frequency (or twice this value), due 
to a hot spot (or spots) in an atmospheric layer of the rotating neutron star.
Many LMXBs (including most of the SXTs) show rich time variability both
at low and at high frequencies, in the form of noise components or quasi
periodic oscillations (QPOs). In particular, QPOs at kilohertz frequencies 
(kHz QPOs), with frequencies ranging from a few hundred Hz up to $1200-1300$ 
Hz (see van der Klis 2000 for a review), have been observed in the emission 
of about 20 LMXBs. Usually two kHz QPO peaks (``twin peaks'') are 
simultaneously observed, the difference between their centroid frequencies 
being in the range 250--350~Hz (usually similar, but not exactly identical,
to the corresponding nearly-coherent frequency of the burst oscillations, 
or half that value). 
%%The kHz QPO frequencies increase when the inferred mass accretion rate 
%%increases. 

The presence and intensity of a magnetic field in LMXBs is an important
question to address.
The widely accepted scenario for the formation of millisecond radio pulsars
is the  recycling of an old  neutron star by a spin-up process  driven by  
accretion of  matter and  angular momentum  from a Keplerian disc, fueled 
{\it via}  Roche lobe overflow of a binary late-type companion (see  
Bhattacharya \& van den  Heuvel 1991 for a review). Once  the accretion  
and spin-up  process ends, the neutron star is visible as a millisecond radio
pulsar. 
The connection between LMXBs and millisecond radio pulsars indicates that
neutron stars in LMXBs have magnetic fields of the order of $B \sim 10^{8}-
10^{9}$ Gauss.  In this case, the accretion disc in LMXBs should be  truncated 
at the magnetosphere,  where   the  disc  pressure  is balanced by the 
magnetic pressure exerted by the neutron star magnetic  field.   
Although widely accepted, there is no direct evidence confirming this 
scenario yet. However, the discovery of coherent X-ray pulsations at 
$\sim 2.5$ ms in \saxj\ (a transient LMXB with an orbital period $P_{\rm orb} 
= 2$ h, Wijnands \& van der Klis 1998) has proved that the neutron star in 
a LMXB can be accelerated to millisecond periods.  
Recently, other two transient LMXBs have been discovered
to be millisecond X-ray pulsars, namely XTE~J1751--305 ($P_{\rm spin} \sim 
2.3$ ms, $P_{\rm orb} = 42$ min, Markwardt et al. 2002) and XTE J0929--314 
($P_{\rm spin} \sim 5.4$ ms, $P_{\rm orb} = 43$ min, Galloway et al. 2002).

Although there are indications for the presence of a (weak) magnetic field
in LMXBs, it is not clear yet whether this magnetic field plays a role in
the accretion process onto the neutron star. If the neutron stars in LMXBs
have magnetic fields and spin rates similar to those of millisecond radio 
pulsars (as implied by the recycling scenario), then the accretion disk 
should be truncated quite far (depending on the accretion rate) from the
stellar surface, and the magnetic field should affect the accretion process.
However, the similarity in the spectral and timing behavior between LMXBs 
containing neutron stars and black hole binaries (see Di Salvo \& Stella 2002, 
van der Klis 2000 for reviews) suggests that the neutron star magnetic field
is so weak (less than $10^8$ Gauss, Kluzniak 1998) that it plays no dynamical 
role, and the disk is truncated quite close to the marginally stable orbit, 
both in neutron star and in black hole systems.

We have proposed a method to constrain the magnetic field of transient LMXBs
containing neutron stars based on their measured luminosity in quiescence
and spin rates (when available, Burderi et al. 2002a).  In this paper we 
apply this method to some SXTs for which the luminosity in quiescence and
the spin period are known.

\section{Constraints on the neutron star magnetic field from the quiescent 
X-ray luminosity}

Burderi et al. (2002a) have shown that it is possible to derive an upper limit 
on the neutron star magnetic field measuring the luminosity of these sources 
in quiescence and comparing it with the expectations from the different 
mechanisms that have been proposed to explain the quiescent X-ray emission of 
neutron star SXTs. In the following we will summarize their conclusions.
There exist three sources of energy which might produce some X-ray luminosity 
in quiescence: 
\begin{itemize}
\item[a)] Residual accretion onto the neutron star surface at very low rate 
(e.g.\ Stella et al. 1994); 
\item[b)] Rotational energy of the neutron star converted into radiation
through the emission from a rotating magnetic dipole, a fraction of
which can be emitted in X-rays (e.g.\ Possenti et al. 2002; 
Campana et al. 1998b, and references therein); 
\item[c)] Thermal energy, stored into the neutron star during previous phases
of accretion, released during quiescence (e.g.\ Brown, Bildsten, \& 
Rutledge 1998; Colpi et al. 2001; Rutledge et al. 2001). 
\end{itemize}
Constraints on the neutron star magnetic field can be derived considering 
processes a and b, in the hypothesis that the neutron star spin frequency is 
known.
Note that, while processes a and b are mutually exclusive, process c will 
probably always contribute to the luminosity in quiescence, reducing the 
amount of emission due to one of the first two processes.

If the neutron star has a non-zero magnetic field, then its magnetospheric
radius, i.e.\ the radius at which the pressure due to the (assumed dipolar) 
neutron star magnetic field equals the ram pressure of the accreting matter, 
can only be inside or outside the light-cylinder radius (i.e.\ the 
radius at which an object corotating with the neutron star, having spin period 
$P$, attains the speed of light $c$, $r_{\rm LC} = c P / 2 \pi$), with 
different consequences on the neutron star behavior. 

If the magnetospheric radius is inside 
the light cylinder radius, scenario a, there should be some matter flow inside 
the light cylinder radius in order to keep the magnetospheric radius small 
enough. Actually, accretion onto a spinning, magnetized neutron star is 
centrifugally inhibited once the magnetospheric radius is outside the 
corotation radius, i.e.\ the radius at which the Keplerian frequency of the 
orbiting matter is equal to the neutron star spin frequency,
%%\begin{equation}
$r_{\rm co} = 1.5 \times 10^6 P_{-3}^{2/3} m^{1/3} \;{\rm cm}$
%%\end{equation}
(where $P_{-3}$ is the spin period in ms and $m$ is the neutron star mass in 
solar masses, $M_\odot$).  In this scenario we have therefore
two possibilities: a1) the magnetospheric radius is inside the co-rotation 
radius, so accretion onto the neutron star surface is possible; 
a2) the magnetospheric radius is outside the co-rotation radius (but still 
inside the light cylinder radius), so the accretion onto the neutron star is 
centrifugally inhibited, but an accretion disk can still be present outside
$r_{\rm co}$ and emit X-rays.

If the magnetospheric radius falls outside the light-cylinder radius, it will 
also be outside the corotation radius. This means that the space surrounding 
the neutron star will be free of matter up to the light cylinder radius. 
It has been demonstrated that a rotating 
magnetic dipole in vacuum emits electromagnetic dipole radiation according to 
the Larmor's formula, and a wind of relativistic particles associated with 
magnetospheric currents along the field lines is expected to arise 
(e.g.\ Goldreich \& Julian 1969). 
Therefore, the neutron star will emit as a radio pulsar.
In this case X-ray emission can be produced by: b1) reprocessing of 
part of the bolometric luminosity of the rotating neutron star into X-rays 
in a shock front between the relativistic pulsar wind and the circumstellar 
matter; b2) the intrinsic emission in X-rays of the radio pulsar.

In all these scenarios we have calculated the expected X-ray luminosity in 
quiescence, which of course depends on the neutron star spin frequency and 
magnetic field (see Burderi et al. 2002a, and references therein, for details). 
This can be compared with the observed quiescent luminosity (which has to 
be considered as an upper limit for the luminosity due to each of these 
processes, given that process c is also expected to contribute) giving an 
upper limit on the magnetic field, once the neutron star spin frequency is 
known. For each of the scenarios above, these upper limits are:
\begin{itemize}
\item[a1)] 
$\mu_{26} \le 0.08 L_{33}^{1/2} m^{1/3} P_{-3}^{7/6}$
\item[a2)] 
$\mu_{26} \le 1.9 L_{33}^{1/2}  m^{-1/4} P_{-3}^{9/4}$
\item[b1)] 
$\mu_{26} \le 0.05 L_{33}^{1/2} P_{-3}^2 \eta^{-1/2}$
\item[b2)] 
$\mu_{26} \le 2.37 L_{33}^{0.38} P_{-3}^2$ 
\end{itemize}
where $\mu_{26}$ is the neutron star magnetic moment in units of $10^{26}$ 
Gauss cm$^3$, $L_{33}$ is the accretion luminosity in units of $10^{33}$ ergs/s, 
and $\eta \sim 0.01-0.1$ is the efficiency in the conversion of the rotational 
energy into X-rays (e.g.\ Campana et al. 1998b; Tavani 1991; Kaspi et al. 1995; 
Grove et al. 1995).  Note that the values assumed for $\eta$ are quite
uncertain given that they are inferred from radio pulsar studies and, of
course, depend on the geometry of the system as well as on the characteristics
of the surrounding environment. However, these values are also in agreement
with the measured quiescent X-ray emission of SXTs (Campana et al. 1998b).

We have already applied this considerations to the case of KS 1731--260, 
getting valuable results (Burderi et al. 2002a).  In the following we
re-calculate the upper limit on the neutron star magnetic field of this
source using the most recent measurement of its quiescence luminosity
($(2-5) \times 10^{32}$ ergs/s in the $0.5-10$ keV energy range) obtained 
from XMM-Newton observations (Wijnands et al. 2002).

KS 1731--260 is a neutron star SXT, which in February 2001
entered a quiescent state after a long period of activity lasted more
than a decade. The quiescent X-ray luminosity of $\sim 10^{33}$ ergs/s was 
measured with BeppoSAX (Burderi et al. 2002a) and Chandra (Wijnands et al. 
2001).  The X-ray spectrum obtained with Chandra is well described by a 
blackbody at a temperature of $\sim 0.3$ keV (Wijnands et al. 2001) or by 
a hydrogen atmosphere model, obtaining an effective temperature of 0.12 keV 
and an emission area radius of $\sim 10$ km (Rutledge et al. 2001).
KS 1731--260 also shows nearly-coherent burst oscillations at 
$\sim 524$ Hz (corresponding to a period of 1.91 ms, which is most 
probably the neutron star spin period, see Muno et al. 2000).  Using this 
information we have applied the method described above to derive constraints 
on the neutron star magnetic field, which turned out to be less than 
$\sim 9 \times 10^{8}$ Gauss (Burderi et al. 2002a).
%% RICALCOLATO USANDO LA LUMINOSITA' IN QUIESCENZA MISURATA CON XMM:
%% 0.5-10 keV X-ray luminosity of approximately (2-5) x 10^32 ergs s-1
%% (Wijnands et al. 2002, ApJ, 573, L45)
A recent XMM-newton observation of this source gave a quiescent luminosity
of $\sim (2-5) \times 10^{32}$ ergs/s, about a factor of two lower than the
previous BeppoSAX and Chandra estimations. Adopting therefore a quiescent
luminosity of $5 \times 10^{32}$ ergs/s and a spin period of 1.9 ms, the 
upper limits to the magnetic field of the neutron star in KS 1731--260 are: 
a1) $\mu_{26} \le 0.2  P_{1.9}^{7/6} m^{1/3}$; 
a2) $\mu_{26} \le 5.9  P_{1.9}^{9/4} m^{-1/4}$;
b1) $\mu_{26} \le 1.3 P_{1.9}^2 \eta_{0.01}^{-1/2}$; 
b2) $\mu_{26} \le 6.6 P_{1.9}^2$. 
Here $P_{1.9}$ is the spin period in units of 1.9 ms, and $\eta_{0.01}$
is the conversion efficiency $\eta$ in units of 0.01.  In any case the 
magnetic field of KS 1731--260 results most probably less than 
$\sim 7 \times 10^8$ Gauss.

\section{Constraints on the magnetic field of \saxj\ and \aql}

SAX J1808.4--3658, another bursting SXT, was the first (low magnetized) LMXB 
to show coherent pulsations, at a frequency of $\sim 401$ Hz, in its persistent 
emission (Wijnands \& van der Klis 1998), thus providing the first direct
evidence of the current evolutionary scenarios according to which LMXBs
are the progenitors of millisecond radio pulsars.  SAX J1808.4--3658 has been
observed in quiescence with XMM: the 0.5--10 keV unabsorbed luminosity was
$5 \times 10^{31}$ ergs/s (Campana et al. 2002).  Using the reasoning 
described above we can calculate the upper limits on the neutron star magnetic
field in this system, which for the various process are respectively:
a1) accretion: 
  $\mu_{26} < 0.054 m^{1/3}$;
a2) propeller:
  $\mu_{26} < 3.4 m^{-1/4}$;
b1) reprocessed radio emission:
  $\mu_{26} < 0.71 \eta_{0.01}^{-1/2}$;
b2) radio pulsar emission:
  $\mu_{26} < 4.69$. 
In this case we find a
maximum neutron star magnetic field strength of $\sim 4.7 \times 10^8$ Gauss. 
A lower limit on the pulsar magnetic field in SAX J1808.4--3658 was
calculated using the observation of coherent pulsations during 
%%absence of centrifugal inhibition of accretion during the decline of 
the 1998 outburst, i.e.\ imposing that the magnetospheric radius is larger
than the neutron star radius at the highest flux level; this gives 
$B \ge 1 \times 10^8$ Gauss (Psaltis \& Chakrabarty 1999).  
Combining this with our upper limit, we can constrain the neutron star 
magnetic field in this system in the rather small range $(1-5) \times 10^8$ 
Gauss. Note that the presence of coherent pulsations in this source excludes
the possibility that the magnetic field is as weak as $\sim 0.05$ Gauss; 
therefore the residual accretion (scenario a1) is excluded in this case. This 
has important consequences as regards the understanding of the origin of the 
quiescent emission. 
In the case of \saxj\ the X-ray spectrum in quiescence is dominated
by the power-law component; the soft blackbody component (if any) contributes 
at least a factor of 15 less than the power law to the quiescent luminosity 
in the 0.5--10 keV range (see Campana et al. 2002). Although the details of 
the emission in the propeller regime are not clear, we would expect that in
this regime the source spectrum is dominated by a thermal component from the 
residual accretion disk that should be present outside the magnetospheric 
radius. This is not the case for \saxj.  Therefore we suggest that while its 
thermal (blackbody) component is almost certainly due to what we called process 
c, i.e.\ cooling of the neutron star heated during the previous accretion 
phase, its non-thermal (power-law) component, which sometimes constitutes most 
of the quiescent emission, is most probably due to dipole emission from the 
radio pulsar (scenarios b1 and/or b2). Note, however, that recent 
observational results on the variability of the soft component during a 
quiescence period of \aql\ pose some problems to the neutron star cooling 
scenario (see Rutledge et al. 2002). 
%%In fact, a residual accretion is excluded by the constraints on the 
%%magnetic field and a propeller scenario would probably result in a 
%%thermal (disk blackbody) spectrum.

\aql\ is a SXT showing type-I X-ray bursts. Based on RXTE/PCA observations 
taken during an outburst in 1997, Zhang et al. (1998) discovered nearly 
coherent oscillations with an asymptotic frequency of 548.9 Hz 
(corresponding to a period of 1.82 ms) during the decay of 
a type I X-ray burst. This signal, as well as similar signals observed during 
type I X-ray bursts from about ten low mass X-ray binaries, likely corresponds 
to the neutron star rotation frequency (or twice its value; for a review see 
Strohmayer 2001). 
\aql\ also shows kHz QPOs: the lower peak frequency varies in the range
$670-930$ Hz, while the upper peak was only marginally detected at 
$\sim 1040$ Hz. The peak separation between the kHz QPOs is $241 \pm 9$ Hz,
inconsistent with (but close to) half the frequency of the burst oscillations.
We will therefore assume that the neutron star in this system is spinning at 
a period of 1.82 ms or 3.64 ms. 
Note, however, that the lack of harmonic content in the burst oscillations 
(see Muno et al. 2002) might suggest that these correspond indeed to the 
neutron star spin frequency. \aql\ was observed several times in quiescence
(with ROSAT, Verbunt et al. 1994; ASCA, Asai et al. 1998; BeppoSAX, Campana
et al. 1998; and Chandra, Rutledge et al. 2001). For the quiescent X-ray 
luminosity of \aql\ we adopt the minimum value reported in the literature,
that is $\sim 1.6 \times 10^{33}$ ergs/s (from Verbunt et al. 1994, 
extrapolated in the 0.5-10 keV energy range and recomputed for a distance 
of 5 kpc, see Rutledge et al. 2002).
Using these parameters we can apply the formulas above to calculate the
upper limit on the neutron star magnetic field in \aql: 
a1) accretion:
  $\mu_{26} < 0.20  P_{1.8}^{7/6} m^{1/3}$;
a2) propeller:
  $\mu_{26} < 9.25  P_{1.8}^{9/4} m^{-1/4}$;
b1) reprocessed radio emission:
  $\mu_{26} < 2.1 P_{1.8}^{2} \eta_{0.01}^{-1/2}$;
b2) radio pulsar emission:
  $\mu_{26} < 7.85 P_{1.8}^{2}$.
Here $P_{1.8}$ is the spin period in units of 1.8 ms.
For a spin period of 1.82 ms, 
the highest magnetic field we get is $\sim 9 \times 10^8$ Gauss. Assuming a
spin period of 3.64 ms, the magnetic field is less constrained, with
a maximum value of $\sim 4 \times 10^9$ Gauss that is obtained in the
propeller (a2) scenario (note that much lower upper limits are obtained
in the case of residual accretion, a1, and reprocessed radio emission, b2).

\section{Discussion and conclusions}

We have applied a method to constrain the magnetic field of transient LMXBs
containing neutron stars based on their measured luminosity in quiescence
and spin rates.  
This gives a magnetic field lower than $\sim 7 \times 10^8$ 
Gauss for KS~1731--260 and lower than $\sim 10^9$ Gauss for \aql, and 
constrains the magnetic field of the millisecond X-ray pulsar \saxj\ in the 
quite narrow range between $10^8$ and $5 \times 10^8$ Gauss.  

In the case of \saxj\ we also find that residual accretion onto the neutron 
star very unlikely contributes to the source luminosity in quiescence, and 
we suggest that the non-thermal (power-law) component of the source spectrum 
in quiescence is probably produced by reprocessed and/or direct dipole emission 
from the radio pulsar, which may switch on at very-low accretion rates.
In the arguments developed above we assume that, once the accretion rate 
significantly decreases during quiescence, the radio pulsar switches on when 
the magnetospheric radius becomes larger than the light cylinder radius. 
In this case, we can estimate the timescale on which the pulsar is expected 
to clean the space up to its light cylinder radius and turn on, which of 
course will depend on the details of the decrease in the mass accretion rate. 
Assuming that the mass accretion rate instantaneously drops to a very low 
value, we have estimated the net force on the disk (considered in the standard 
Shakura \& Sunyaev, 1973, configuration) induced by the magnetic field 
pressure. This rough calculation gives a timescale of $\sim 1-10$ s (depending 
on the residual accretion rate and on the viscosity parameter $\alpha$) in the 
case of \saxj.

However, more evidences are needed to confirm this suggestion, as, for 
instance, the detection of pulsed radio emission in quiescence. Indeed, 
despite thoroughly searched in radio during its X-ray quiescent phase, 
no pulsed radio emission has been detected from SAX J1808.4-3658 up to now
(see e.g.\ Burgay et al. 2002). This can be caused by the 
presence of a strong wind of matter emanating from the system: the mass 
released by the companion star swept away by the radiation pressure of the 
pulsar, as predicted in the so-called radio-ejection model (Burderi et al.
2001; see also Burderi et al. 2002b).  This means that SAX J1808.4-3658 may 
show radio pulsations in quiescence when observed at frequencies higher than 
the standard 1.4 GHz (the frequency at which radio pulsars are normally 
searched), where the free-free absorption is less severe.  

We note, however, that our magnetic field upper limits are subject to 
significant uncertainties (that will be probably addressed by future studies)
due to our lack of understanding of the structure of the accretion flow in 
quiescence.  In particular, our calculations consider an accretion-disk 
geometry.  In a different geometry, it is not clear whether a clear 
distinction between the residual accretion scenario and the propeller 
scenario can be made. Menou et al. (1999) have presented a propeller scenario 
that still allows for partial residual accretion if accretion in quiescent 
neutron star SXTs occurs via a quasi-spherical Advection Dominated Accretion 
Flow (ADAF), rather than a thin disk. This allows some material to accrete 
near the poles, bypassing the centrifugal barrier (see also Zhang et al. 1998). 
However, only a very small fraction (if any) of the total accretion rate in 
quiescence is expected to accrete onto the neutron star surface (Menou et al. 
1999), consistently with the fact that no coherent pulsations are detected 
during quiescence. Other significant uncertainties are in the assumed values
of the efficiency $\eta$ in the conversion of the pulsar spin-down energy 
into X-rays in the shock front, b1, scenario and in the fraction of the 
intrinsic radio-pulsar emission emitted in X-rays (scenario b2).

Although subject to some uncertainties our upper limits on the neutron 
star magnetic field are reasonable and in agreement with limits found from
different considerations.
The presence of a weak, but not negligible, magnetic field in LMXBs 
has been invoked to explain some observational facts such as the QPO at 
$\sim 20-60$ Hz (the so-called low-frequency QPO in atoll sources or
horizontal branch oscillations, HBOs, in Z sources; Psaltis et al. 1999),
or the disappearance of the kHz QPOs at low and high inferred mass accretion
rates (e.g.\ Campana 2000; Cui 2000). In particular, linking the kHz QPO 
observability to variations of the neutron star magnetospheric radius, in 
response to changes in the mass accretion rate, Campana (2000) estimates
a magnetic field of $B \sim (0.3-1) \times 10^8$ Gauss for Aql X-1 and of
$B \sim (1-8) \times 10^8$ Gauss for Cyg X-2. 
A method for determining the B-field around neutron stars based on observed 
kilohertz and other QPOs frequencies, in the framework of the transition
layer QPO model (Titarchuk, Lapidus, \& Muslimov 1998), gives dipole fields 
with the strengths of $10^7-10^8$ Gauss on the neutron star surface for 
4U 1728--34, GX 340+0, and Scorpius X-1 (Titarchuk, Bradshaw, \& 
Wood 2001).

The accurate measurement of the luminosity in quiescence 
of other SXTs (and in particular of the other X-ray millisecond pulsars, for 
which the spin period is precisely determined), certainly possible with the 
high sensitivity of the instruments on board Chandra and XMM-Newton, will 
give important information about the magnetic field in these systems and 
therefore about the connection between the populations of LMXBs and 
millisecond radio pulsars as well as about the influence of the magnetic field
in the accretion process onto the neutron star.

\begin{acknowledgements}
This work was performed in the context of the research network "Accretion 
onto black holes, compact stars and protostars", funded by the European 
Commission under contract number ERB-FMRX-CT98-0195, and was partially 
supported by the Netherlands Organization for Scientific Research (NWO). 
LB thanks MIUR for financial support.
\end{acknowledgements}

\end{document}